\documentclass[12pt,a4paper]{article}
\usepackage[applemac]{inputenc}
\usepackage[T1]{fontenc}
\usepackage{graphicx}
\usepackage{amsmath, amsfonts} 		
\usepackage{units, upgreek} 			
\usepackage[comma, sort&compress]{natbib}

\usepackage[margin=30pt, font=small,labelfont=bf]{caption}	

\usepackage{geometry}
\title{Estimating Moving Average Processes \\with an improved version of Durbin's Method}
\author{Maximilian Ludwig\footnote{University of Hamburg, Department of Economics, \texttt{mludwig@mludwig.org}. The latest version of this paper and a ready to use MatLab implementation of the estimator are available at \texttt{http://mludwig.org/research.html}.}}
\date{this version: \today, \\ \small initial version: April 12, 2013}

\begin{document}
\maketitle
\abstract{
\footnotesize
This paper provides a simple method to estimate both univariate and multivariate MA processes. Similar to Durbin's method, it rests on the recursive relation between the parameters of the MA process and those of its AR representation. This recursive relation is shown to be valid both for invertible / stable and non invertible / unstable processes under the assumption that the process has no constant and started from zero. This makes the method suitable for unit root processes, too.}

\section{Introduction and Summary}
A classic method to estimate the parameters of pure MA processes is provided by \citet{Durbin1959}. This method is a two step estimation. First, an AR process of large order is estimated using OLS. The parameter estimates are then used to estimate the parameters of the MA process using the Yule-Walker estimator, which is shown to be efficient for this purpose by Durbin. 

In this paper, I show that under one assumption, the well know recursive formulae to compute the MA representation of an AR process / the AR representation of an MA process can be derived without assuming stability of the AR / invertibility of the MA  process. Inspecting the recursive formula reveals that the AR parameters can themselves be viewed as generated by an AR process parameterized by the MA. Therefore, Durbin's method is basically a Yule-Walker estimation of the latter AR($q$) process on parameters from the initial AR estimation. It is well known and documented by e.g. \cite{SandStoi2006} that the precision of estimates obtained by Durbin's method is unsatisfactory if one or more roots of the data generating MA process are close to the unit circle. The reason for this is simple: As documented by \citet{TjPaul1983}  and others, the Yule-Walker estimator is not well suited to estimate AR processes with roots close to the unit circle and biased in case one root is on the unit circle. Since the OLS estimator is consistent also for AR processes with unit root, a natural modification of Durbin's method is the use of the OLS estimator for the second estimation. But since the Yule-Walker estimator is efficient, one faces a trade-off between a lower efficiency for MA processes with roots away from the unit circle and more reliable estimates in case of one or more roots close to unity. However, by exploiting the structural information implied by the recursive formula one can use a restricted OLS estimator to counteract this disadvantage. I perform a simulation for the parameter space of an MA(2) process which indicates that this approach yields substantial gains in efficiency in comparison to Durbin's method for a very large part of the investigated parameter space. Finally, a generalization of the approach for multivariate MA processes is provided.

\section{The recursive formulae}
\subsection{A more general MA representation}
The most important tool throughout the analysis is the so called companion form of an
AR($p$) process. This means that the process is noted as VAR(1):
\begin{align}\label{Basic}
\underbrace{
\begin{bmatrix}
y_{t} \\
 y_{t-1}\\
 \vdots \\
 y_{t-p+1}
\end{bmatrix}
}_{\rm{y_{t}}} 
=
\underbrace{
\begin{bmatrix}
v			\\
0			\\
\vdots		\\
0			\\
\end{bmatrix}}_{\rm v}
+
\underbrace{
\begin{bmatrix}
\phi_1  	& \phi_2  	& 	\dots	& \phi_p  \\
 1 		&   	0		& 	\dots	& 0		\\
\vdots	&	\ddots	& 			& \vdots	\\
 0		& 			&  1			& 0\\
\end{bmatrix}
}_{\rm F} 
\underbrace{
\begin{bmatrix}
y_{t-1} \\
 y_{t-2}\\
 \vdots \\
 y_{t-p}
\end{bmatrix}
}_{\rm y_{t-1}}
+
\underbrace{
\begin{bmatrix}
\varepsilon_{t}  \\
0 \\
\vdots \\
0 
\end{bmatrix}
}_{\rm \upvarepsilon_{t} }  
\end{align}
For a given vector of starting values ${\rm y_{t-\tau}}$ it is always possible to write the process as a function of errors terms, starting values and time:
\begin{align*}
\rm y_{t-\tau +1} = & \rm v + F y_{t - \tau} + \upvarepsilon_{t-\tau +1}
\\
\rm y_{t-\tau +2} = & \rm v + F v +  F  F y_{t - \tau} 
+ F \upvarepsilon_{t-\tau +1} 
+\upvarepsilon_{t-\tau +2}
\\
\rm y_{t-\tau +3} = & \rm v + F v + F^2 v + F  F^2 y_{t - \tau} 
+ F^2 \upvarepsilon_{t-\tau +1} 
+ F \upvarepsilon_{t-\tau +2}
+ \upvarepsilon_{t-\tau +2}
\\
& \vdots
\\
{\rm y_t }
= & 
\Big( {\rm I} + \sum_{j=1}^{\tau -1}{ \rm  F^j \Big) v + F^\tau y_{t-\tau}} 
+ \sum_{j=1}^{\tau - 1} \rm F^j \upvarepsilon_{t-j} + \upvarepsilon_{t}
\end{align*}
Defining $\rm F^0 \equiv I$ and using $\rm (I - F)^{-1}(I - F) ( \sum_{j=0}^{\tau-1} \rm F^j ) = (I - F)^{-1}(I- F^\tau) $ one can write this equation in a manner that I call the \emph{generalized moving average representation}:
\begin{align}\label{genMA}
{\rm y_t} = \underbrace{{\rm (I - F)^{-1}(I - F^\tau) v + F^\tau y_{t-\tau} }}_{\rm w_t}
+ \sum_{j=0}^{\tau-1} \rm F^j \upvarepsilon_{t-j} 
\end{align}
This name seems appropriate since the generalized moving average representation contains the textbook moving average representation in its companion form version as a special case. This is easy to see in case of distinct eigenvalues of $\rm F$. It is well known that in this case, $\rm F$ can be decomposed as $\rm W \Lambda W^{-1}$ where $\Lambda$ is a diagonal matrix of the eigenvalues and W a matrix containing the eigenvectors. This allows to write $\rm F^\tau  = W \Lambda^\tau W^{-1}$. As shown by \citet[][pp.21]{Hamilton} the eigenvalues of F are equivalent to the inverse roots of the lag polynomial and hence the process is stable if the norm of all eigenvalues of F is smaller than one. Then for $\tau \rightarrow \infty \quad \rm \Lambda^\tau\rightarrow O_p$ where $\rm O_p$ denotes a matrix of zeros in $\mathbb{R}^{p \times p}$. Thus: $\lim_{\tau \to \infty} {\rm E [ w_t]} = \rm (I - F)^{-1}v = \upmu$, 
%
%
which allows to write \eqref{genMA} for $\tau=\infty$ and eigenvalues in the unit circle as:
\begin{align}\label{TextBookMA}
{\rm y_t} = \upmu
+ \sum_{j=0}^{\infty} \rm F^j \upvarepsilon_{t-j} 
\end{align}
The literature rarely approaches AR processes with the companion form but uses lag polynomials. However, their use in case one seeks to invert a lag  polynomial is linked to the assumption of $\tau=\infty$ and stability / invertibility. For the important special of a process with $v=0$, these assumptions can be replaced with a single assumption that  might be more sensible in some settings:

\medskip
\noindent
\textbf{Assumption:} \emph{The ''memory'' of the process has been emptied at some time in the past, meaning $\rm y_{t-\tau} = 0$.}

\medskip
\noindent
Under this assumption $\rm w_t=0$ for any parametrization and \eqref{genMA} looks quite similar to \eqref{TextBookMA} with $\upmu=0$:
\begin{align}\label{genMAv0}
{\rm y_{t}}=\sum_{j=0}^{\tau-1} \rm F^j \upvarepsilon_{t-j} 
\end{align}
However, \eqref{genMAv0} is an MA representation of both stable and unstable AR processes. From here on I shall impose both $v=0$ and $\rm y_{t-\tau} = 0$, where $\tau$ might be both finite or infinite.

\subsection{The recursive formula for AR processes}

To my best knowledge, the only way the recursive formula has been derived previously is by using the lag polynomial of the AR process, that is: assuming $\tau = \infty$ and stability. I shall reproduce this approach here for convenience. Consider an  AR($p$) process without constant in lag notion: $y_t \phi(L)= \varepsilon_{t} $. Noting $\phi(L)^{-1} \equiv \psi(L)$ gives the identity:
\begin{align*}
1 = &\phi(L) \psi(L)
=  (1 - \sum_{j=1}^p \phi_j L^j)(\sum_{i=0}^\infty \psi_i L^i)
\\
= & (\sum_{i=0}^\infty \psi_i L^i) - \phi_1 L (\sum_{i=0}^\infty \psi_i L^i) - \dots - \phi_p L^p (\sum_{i=0}^\infty \psi_i L^i)
\\
= & \psi_0 + (\psi_1 - \phi_1 \psi_0) L + (\psi_2 - \phi_1 \psi_1 - \phi_2 \psi_0) L^2 + \dots
\\
& \dots + (\psi_p - \phi_1 \psi_{p-1} - \phi_2 \psi_{p-2} - \dots \phi_{p} \psi_0) L^p \dots
\\  & \dots
+ (\psi_{p+1} - \phi_1 \psi_p - \dots - \phi_p \psi_1) L^{p+1} + \dots
\end{align*}
Since $\psi_0 \equiv 1$ the parameters of the MA representation can be recursively computed as $\psi_1 =  \phi_1$, $\psi_2 =   \phi_1 \psi_1  + \phi_2$ and so on.

To derive the same recursive formula from \eqref{genMAv0} one has to compute the powers of F, which seems harder than it actually is. Consider $\rm F^2$:
\begin{align}  \label{FSq}
\left[
\begin{smallmatrix}
\phi_1  	&	\dots	& 	\dots	&	\phi_p  	\\
 1 		&   	0		&	\dots	& 	0		\\
 \vdots	&	\ddots	& 			& 	\vdots	\\
 0		& 	0		& 	1		& 	0		\\
\end{smallmatrix}
\right]
\left[
\begin{smallmatrix}
\phi_1  	&	\dots	& 	\dots	&	\phi_p  	\\
 1 		&   	0		&	\dots	& 	0		\\
 \vdots	&	\ddots	& 			& 	\vdots	\\
 0		& 	0		& 	1		& 	0		\\
\end{smallmatrix}
\right]
= 
\left[
\begin{smallmatrix}
\phi_1^2+ \phi_2	& \dots 	&  \phi_1 \phi_{p-2} + \phi_{p-1}	&	\phi_1 \phi_{p-1} + \phi_p &	\phi_1 \phi_{p} \\
\phi_1			& \dots 	&	\phi_{p-2}				&	\phi_{p-1}			&	\phi_p		\\
1				&\dots	&	0						&	0					&	0	\\
0				& \ddots	& 							&	0					&	0\\
\vdots			&		&							&	\vdots				&	\vdots\\
0				&		& 1							&	0					&	0\\
\end{smallmatrix}
\right]
\end{align}
Obviously, the first column contains the first parameters of the MA representation. Since all elements of $\rm \upvarepsilon_{t-j}$ except the first are zero, one can ignore the values in all columns of ${\rm F}^j$ except for the first one.  Knowing $\rm F^2$ and recalling that $\rm F^{j}$ is generated by the multiplication ${\rm F} \times {\rm F}^{j-1} $ allows to compute $\rm F^3$ as:
\begin{align*}
\rm F^3 =  &
\left[
\begin{smallmatrix}
\phi_1  	&	\dots	& 	\dots	&	\phi_p  	\\
 1 		&   	0		&	\dots	& 	0		\\
 \vdots	&	\ddots	& 			& 	\vdots	\\
 0		& 	0		& 	1		& 	0		\\
\end{smallmatrix}
\right]
\left[
\begin{smallmatrix}
\psi_2	&	\varpi  & \dots  & \varpi \\
\psi_1	&	\varpi  & \dots  & \varpi \\
1		&	\varpi  & \dots  & \varpi \\
0		&	\varpi  & \dots  & \varpi \\
\vdots	&	\vdots	&			&	\vdots	\\
0		&	\varpi  & \dots  & \varpi \\
\end{smallmatrix}
\right]
=
\left[
\begin{smallmatrix}
\phi_1 \psi_2 + \phi_2 \psi_1 + \phi_3	&	\varpi  & \dots  & \varpi \\
\psi_2							&	\varpi  & \dots  & \varpi \\
\psi_1							&	\varpi  & \dots  & \varpi \\
1								&	\varpi  & \dots  & \varpi \\
0								&	\varpi  & \dots  & \varpi \\
\vdots							&	\vdots	&		&	\vdots	\\
0								&	\varpi  & \dots  & \varpi \\
\end{smallmatrix}
\right]
\end{align*}
where $\varpi$ indicates values that are of no interest for the analysis. Thus, $\psi_3 = \phi_1 \psi_2 + \phi_2 \psi_1 + \phi_3$  and obviously one can compute $\rm F^4$ and all ${\rm F^j}$ for $ j=2, \dots, \tau$ in a similar manner. Note that for $j > p$ one gets:
\begin{align*}
\rm F^j = 
\left[
\begin{smallmatrix}
\phi_1  	&	\dots	& 	\dots	&	\phi_p  	\\
 1 		&   	0		&	\dots	& 	0		\\
 \vdots	&	\ddots	& 			& 	\vdots	\\
 0		& 	0		& 	1		& 	0		\\
\end{smallmatrix}
\right]
\left[
\begin{smallmatrix}
\psi_{j-1}	&	\varpi  & \dots  & \varpi \\
\psi_{j-2}	&	\varpi  & \dots  & \varpi \\
\vdots	&	\vdots	&	&	\vdots	\\
\psi_{j-p}	&	\varpi  & \dots  & \varpi \\
\end{smallmatrix}
\right]
\end{align*}
Which means that for $j > p$, the MA parameters are generated by the difference equation $\psi_j = \sum_{i=1}^p \phi_i \psi_{j-i}$.

\subsection{The recursive formula for MA processes}

Using the Lag operator, a zero mean MA($q$) process can be noted as
$ y_t =\psi(L) \varepsilon_{t} $. The corresponding AR($\infty$) process $\phi(L) y_t= \varepsilon_{t} $ can be used in the same way as above to gain the identity
\begin{align*}
1 = & \psi(L) \phi(L)
= \phi(L) + \psi_1 L \phi(L) + \dots + \psi_q L^q \phi(L)
\\
= & (1 - \sum_{j=1}^\infty  \phi_j L^j) + \psi_1 L (1 - \sum_{j=1}^\infty  \phi_j L^j)  + \dots
\\
\Leftrightarrow \;
0 = & - \sum_{j=1}^\infty  \phi_j L^j  + \psi_1 L - \sum_{j=1}^\infty  \psi_1 \phi_j L^{j+1}
+ \psi_2 L^2 - \sum_{j=1}^\infty  \psi_2 \phi_j L^{j+2} + \dots
\\
\Leftrightarrow \;
0 = & (\psi_1 - \phi_1) L + (\psi_2 - \phi_2 - \psi_1 \phi_1) L^2 + \dots
\\
 & \dots +(\psi_q - \phi_q - \psi_1 \phi_{q-1} - \psi_2 \phi_{q-2} - \dots - \psi_{q-1} \phi_1) L^q \dots
\\ 
& \dots + (- \phi_{q+1} - \psi_1 \phi_{q} - \dots - \psi_q \phi_1) L^{q+1} \dots
\\ 
& \dots + (- \phi_{q+2} - \psi_1 \phi_{q+1} - \dots - \psi_q \phi_{2}) L^{q+2} + \dots
\end{align*}
Thus, $\phi_1 = \psi_1$, $\psi_2 = \phi_2 + \psi_1 \phi_1$ and so on. To show that this is indeed the AR representation of both invertible and noninvertible MA processes given $v=0$ and ${\rm y_{t-\tau}}=0$, one needs to note \eqref{genMAv0} compatible to the companion form of an AR($p$) process. Suppose we have $T> q$ observations of $y_t$, know that they were generated by an MA($q$) process and look for an AR($l$) representation of this process. To do so, write down the structure of the data generating process in a -- sort of -- companion form for MA processes.
\begin{align*}
\underbrace{
\left[
\begin{smallmatrix}
y_t		\\
y_{t-1}	\\
\vdots	\\
y_{t-l}
\end{smallmatrix}
\right]
}_{\rm y_t}
 - 
 \underbrace{
\left[
\begin{smallmatrix}
\varepsilon_{t}	\\
0			\\	
\vdots		\\
0			
\end{smallmatrix}
\right]
}_{\rm \upvarepsilon_{t}}
=
&
\underbrace{
\left[
\begin{smallmatrix}
\psi_1	&	\varpi  & \dots  & \varpi \\
1		&	\varpi  & \dots  & \varpi \\
\vdots	&	\vdots	&			&	\vdots	\\
0		&	\varpi  & \dots  & \varpi \\
\end{smallmatrix}
\right]
}_{\rm M_1}
\left[
\begin{smallmatrix}
\varepsilon_{t-1}	\\	
0			\\
\vdots		\\
0			\\
\end{smallmatrix}
\right]
+
\underbrace{
\left[
\begin{smallmatrix}
\psi_2	&	\varpi  & \dots  & \varpi \\
\psi_1	&	\varpi  & \dots  & \varpi \\
1		&	\varpi  & \dots  & \varpi \\
0		&	\varpi  & \dots  & \varpi \\
\vdots	&	\vdots	&			&	\vdots	\\
0		&	\varpi  & \dots  & \varpi \\
\end{smallmatrix}
\right]
}_{\rm M_2}
\left[
\begin{smallmatrix}
\varepsilon_{t-2}	\\	
0			\\
\vdots		\\
0			\\
\end{smallmatrix}
\right]
\dots
\\
\dots + & 
\underbrace{
\left[
\begin{smallmatrix}
\psi_q	&	\varpi  & \dots  & \varpi \\
\vdots	&	\vdots	&	&	\vdots	\\
1		&	\varpi  & \dots  & \varpi \\
0		&	\varpi  & \dots  & \varpi \\
\vdots	&	\vdots	&	&	\vdots	\\
0		&	\varpi  & \dots  & \varpi \\
\end{smallmatrix}
\right]
}_{\rm M_q}
\left[
\begin{smallmatrix}
\varepsilon_{t-q}	\\	
0			\\
\vdots		\\
0			\\
\end{smallmatrix}
\right]
+
\underbrace{
\left[
\begin{smallmatrix}
0		&	\varpi  & \dots  & \varpi \\
\psi_q	&	\varpi  & \dots  & \varpi \\
\vdots	&	\vdots	&			&	\vdots	\\
1		&	\varpi  & \dots  & \varpi \\
0		&	\varpi  & \dots  & \varpi \\
\vdots	&	\vdots	&	&	\vdots	\\
0		&	\varpi  & \dots  & \varpi \\
\end{smallmatrix}
\right]
}_{\rm M_{q+1}}
\left[
\begin{smallmatrix}
\varepsilon_{t-q-1}	\\	
0			\\
\vdots		\\
0			\\
\end{smallmatrix}
\right]
\dots
\\
\dots +
&
\underbrace{
\left[
\begin{smallmatrix}
0		&	\varpi  & \dots  & \varpi \\
\vdots	&	\vdots	&			&	\vdots	\\
0		&	\varpi  & \dots  & \varpi \\
\psi_q	&	\varpi  & \dots  & \varpi \\
\end{smallmatrix}
\right]
}_{\rm M_{q+l}}
\left[
\begin{smallmatrix}
\varepsilon_{t-l-q}	\\	
0			\\
\vdots		\\
0			\\
\end{smallmatrix}
\right]
\end{align*}
where  $\varpi$ denotes again values of no interest for the analysis and ${\rm M_i} \in \mathbb{R}^{l \times l}$. To find the recursive formula, use the knowledge that this MA process must be the generalized MA representation of the AR process we are interested in. Since for the AR process we are looking for $\rm F \upvarepsilon_{t-1} = M_1\upvarepsilon_{t-1}$ it is obvious that $\psi_1 = \phi_1$. The first column of $\rm M_2$ has been computed as the first column of $\rm F^2$ in \eqref{FSq}. Note that one can also think about the computation of the first column of $\rm M_2$ as computing the first column of $\rm F M_1$. This means that one can compute the first column of $\rm M_3$ by using the first column of $\rm F M_2$, which has the structure:
\begin{align*}
\rm F^3 = 
\left[
\begin{smallmatrix}
\phi_1  	&	\dots	& 	\dots	&	\phi_p  	\\
 1 		&   	0		&	\dots	& 	0		\\
 \vdots	&	\ddots	& 			& 	\vdots	\\
 0		& 	0		& 	1		& 	0		\\
\end{smallmatrix}
\right]
\left[
\begin{smallmatrix}
\psi_2	&	\varpi  	& \dots  	& \varpi \\
\psi_1	&	\varpi  	& \dots  	& \varpi \\
1		&	\varpi  	& \dots  	& \varpi \\
0		&	\varpi  	& \dots  	& \varpi \\
\vdots	&	\vdots	&		&	\vdots	\\
0		&	\varpi  	& \dots  	& \varpi \\
\end{smallmatrix}
\right]
\end{align*}
thus, $\psi_3 = \phi_1 \psi_2  + \phi_2 \psi_1 + \phi_3$. 
Continuing with this approach, one finds
\begin{align*}
\psi_4 = & \phi_1 \psi_3 + \phi_2 \psi_2 + \phi_3 \psi_1 + \phi_4
\\
\psi_5 = & \phi_1 \psi_4 + \phi_2 \psi_3 + \phi_3 \psi_2 + \phi_4 \psi_1 + \phi_5
\\
& \vdots
\end{align*}
As soon as $\psi_{q}$ is reached, the structure is:
\begin{align*}
0 = & \phi_1 \psi_q + \dots + \phi_{q}\psi_1 + \phi_{q+1}
\\
0 = & \phi_2 \psi_q + \dots + \phi_{q+1}\psi_1 + \phi_{q+2}
\\
& \vdots
\end{align*}
This recursive relation resembles to one derived by using the lag-operator and assuming invertibility / $\tau = \infty$.

\section{Improving Durbin's Method}

I just demonstrated that for $j \geq q$, the $j$th parameter of the AR representation of an MA($q$) process is generated by a $q$th order difference equation:
\begin{align}
\nonumber
0=\phi_{j-q} \psi_q + \dots + \phi_{j-1}\psi_1 + \phi_{j}
\;\Leftrightarrow\;
\label{THE_FINDING}
\phi_j = - \sum_{i=1}^q \psi_{i}\phi_{j-i} 
\end{align}
As noted in the introduction, the method of \citet{Durbin1959} can be viewed as estimating this equation, viz estimating an AR on the parameters of the initial AR estimate. Note that the recursive formula implies that if $j\leq q$, the $j^{\text{th}}$ parameter of the initial AR is not generated by the full difference equation. Therefore, using up to the $q^{\text{th}}$ parameter of the initial AR for the second estimation will usually not improve the quality of the estimate since these observations are not generated by the process one seeks to estimate. The need to drop those estimates makes the usage of the information that $\phi_1  = \psi_1$ particularly helpful because the first parameters are often the ones estimated with the greatest accuracy. 

A classic approach to include stochastic a priori information into OLS estimation is the $f$-class estimator provided by \cite{Theil1963}. Note the a prioi information as
\begin{align*}
\underbrace{
\begin{bmatrix}
1 	&  0 & \dots & 0 \\
\end{bmatrix}}_{\rm R}\uppsi 
= \hat{\phi}_1  + \epsilon
\end{align*}
Where $\epsilon$ denotes the estimation error from using $\hat{\phi}_1$ instead of  the true parameter. Denoting the variance of $\epsilon$ as $\sigma^2(\hat{\phi_i})$, the $f$-class estimator for $\uppsi$ is:
\begin{align*}
\rm 
\hat{\uppsi}^r = -
\Bigg(\frac{1}{\sigma^2}X'X + R'\frac{1}{\sigma^2(\hat{\phi_i})}R\Bigg)^{-1}
\Bigg(\frac{1}{\sigma^2}X'y + R'\frac{1}{\sigma^2(\hat{\phi_i})} 
\big(-\hat{\phi}_1\big)\Bigg)
\end{align*}
where X denotes matrix of explanatory variables, y is the vector containing the depended variables and $\sigma^2$ is the variance of the disturbances. As usual, one has to replace $\sigma^2$ and $\sigma^2(\hat{\phi_i})$ with their OLS estimates $\hat{\sigma}^2$ and $\hat{\sigma}^2(\hat{\phi_i})$. The asymptotic variance of the estimates is given by the diagonal elements of 
\begin{align*}
\rm \Bigg(\frac{1}{\sigma^2}X'X + R'\frac{1}{\sigma^2(\hat{\phi_i})}R\Bigg)^{-1}
\end{align*}
Note that this gives a second reason for the exclusion of $\hat{\phi}_1$ from X and y: it ensures that the assumption of no correlation between $\epsilon$ and the innovations of the core regression model, which underlies the $f$-class estimator, is not violated.  

\section{Some remarks and a simulation}

The recursive formula is helpful to think about the problems concerning MA processes with roots in the unit circle. It implies that such a process has an AR representation whose parameters usually explode for $p \rightarrow \infty$.  At the same time there exists a ''sibling'' MA process with the same first and second moment and all roots outside of the unit circle, implying a non explosive AR representation.  Requiring $\rm w_t=0$ makes sure that the first and second moment are stationary, which tends to make an estimator pick the representation with non explosive parameters and thus the invertible MA process.  This means that there is no hope in estimating the parameters of a noninvertible MA process with the help of the recursive formula by using some kind of restricted estimator.   Since the parametrization one tries to estimate is unlikely, the estimator will use the remaining degrees of freedom to yield an MA process that is ''as invertible as possible'' given the restrictions and thus biased estimates. 

That said, one should also appreciate the relative scale of the problem. Consider an MA process with one root that is just a little smaller than one, say $0.98$. If a practitioner knows this from theory and is confronted with only a few hundred observations, he or she should no spend sleepless nights worrying about the bias induced by the noninvertibility. The reason is simple: the invertible sibling of the MA has the same roots except one that is  $\nicefrac{1}{0.98}$ instead of $0.98$. Therefore its parameters are often very similar. In fact, estimates for an MA process with roots outside the unit circle except one might be as precise than the estimates for a process with the smallest root of, say, $1.001$. The reason is as follows: The AR representation of an invertible MA process is finite if one takes the precision of a digital computer as zero. However, such an AR representation of an MA process with a root of approximately one is very very large. This means that faced with a few hundred observations generated by a close-to unit root MA process, one can not avoid to specify an insufficient number of AR lags which distorts the parameter estimates. A reasonable AR representation of the invertible counterpart of an MA process with 0.98 as smallest root tends to be smaller and thus induces a smaller truncation bias. Therefore, the effects of the smaller truncation bias can counterbalance the general bias introduced by estimating the ''wrong'' process if a noninvertible MA process has only one root inside but close to the unit circle.

To illustrate this point as well as the gains of the using a restriced OLS estimator instead of Durbin's original estimator, I simulated both estimation procedures in the parameter space of an MA(2) process. I investigate the region $\psi_1, \psi_2 \in [-2.2, +2.2]$ which is divided into a grid with 221 points and partition the simulations for invertible an noninvertible processes. To study the gain when estimating invertible processes, all points in the grid that yield noninvertible processes are skipped. To study the performance when estimating noinvertible processes, points where either no root is in the unit circle or where the norm of the smallest root falls short of 0.8 are skipped. 

\begin{figure}[htbp]
\begin{center}
\includegraphics[scale=0.55]{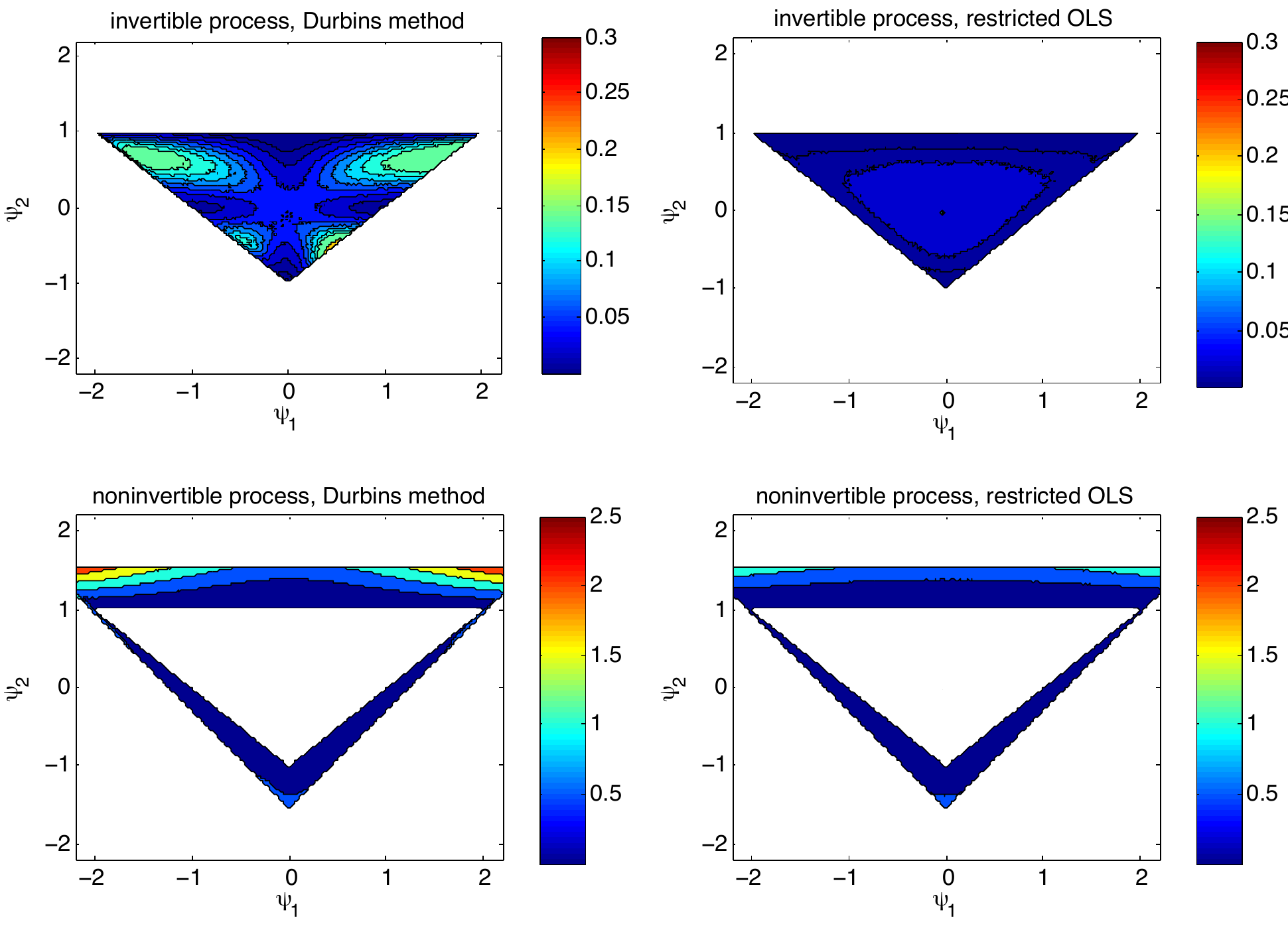}
\caption{Simulation for the parameter space of an MA(2) process. The lower panel exhibits non invertible processes with roots between 0.8 and 1.}\label{SimRes}
\end{center}
\end{figure}

For each point in the grid that is not skipped, 400 artificial observations with standard normal innovations are generated. An AR(100) is estimated using OLS and these estimates are used to estimate an AR(2) process with the Yule-Walker estimator and without dropping any estimate (Durbin's Method) and with Theil's $f$-class estimator using the restriction $\hat{\psi}_1 = \hat{\phi}_1$ while dropping $ \hat{\phi}_1$ and $ \hat{\phi}_2$ from the sample. The squared difference between the true parameters and estimates are computed and saved, after repeating this procedure 500 times for the particular point in the grid the mean of all squared differences is taken as estimate of the MSE. Figure \ref{SimRes} exhibits the results. 

Note that in figure \ref{SimRes},  there are small regions where Durbin's methods yields slightly better results than restricted OLS. However, restricted OLS yields a far smaller MSE for wide areas of the parameter space. Further note that, as stated above, the MSE tends to increase quite moderate outside of the invertibility triangle.

A question of great practical importance is of what order the initial AR estimation should be. This question is tedious since it is a trade off: on the one hand, a larger order of the AR allows the sample which is used to estimate the MA process to be larger and tends to reduce the truncation bias. On the other hand, the precision of the estimates is the worse the more parameters have to be estimated. There is some fairly elaborate work on the question of the optimal trade-off by \citet{Bro2000}. However, in most cases a simple rule of thumb works satisfactory well: make the AR process as large as possible but avoid less than 4 data points per estimate.

\section{Generalization for Multivariate Processes} 

Denoting the parameters of a $k$-dimensional MA process as $\Psi_j$ and the parameters of its VAR representation as $\Phi_i$, it is easy to see that exactly the same recursive formulae apply for multivariate processes, given the assumption applies for each of the time series and the constants are zero. In this case one can express $\Phi_m$ with $m > q$ as:
\begin{align*}
\Phi_m = - \sum_{j=1}^q \Phi_{m-j} \Psi_j 
\end{align*} 
Which means that, given $T >> q$ estimates for the parameters of the VAR, the data model for the estimation of a multivariate MA process is
\begin{align*}
\underbrace{\begin{bmatrix}
 \hat{\Phi}_{T}		\\
 \hat{\Phi}_{T-1}	\\
\vdots		\\
 \hat{\Phi}_{2q+1}		\\
\end{bmatrix}}_{\tilde{\bf{ \Phi}}}
 =  
-
\underbrace{
\begin{bmatrix}
 \hat{\Phi}_{T-1}	& 	 \hat{\Phi}_{T-2}	& 	\dots		& 	 \hat{\Phi}_{T-q}	\\
 \hat{\Phi}_{T-2}	& 	 \hat{\Phi}_{T-3}	& 	\dots		& 	 \hat{\Phi}_{T-q-1}	\\
\vdots		&	\vdots		&			&	\vdots		\\
 \hat{\Phi}_{2q}	 	& 	 \hat{\Phi}_{2q-1}		& 	\dots		& 	 \hat{\Phi}_{q+1}		\\
\end{bmatrix}}_{{\bf \Phi} \in \mathbb{R}^{k(T-2q-1)\times k q}}
\underbrace{
\begin{bmatrix}
\Psi_1 	\\	\Psi_2	\\	 \vdots	\\	 \Psi_q	\\
\end{bmatrix}}_{\tilde{\bf{\Psi}}}
+ {\rm u}
\end{align*}
It is easy to see that one can divide this into $k$ single standard regression models of the kind $\upphi_j = - {\bf \Phi} \uppsi_j + {\rm u}_j$, where $\upphi_j$ is the $j$th column of $\tilde{\bf{\Phi}}$ and $\uppsi_j$ is the $j$th column of $\tilde{\bf{\Psi}}$, implying that the unrestricted OLS estimate is
\begin{align*}
\hat{\uppsi}_j = -\big( {\bf \Phi' \Phi} \big)^{-1} {\bf \Phi}' \upphi_j
\end{align*}
However, for multivariate processes it is also very useful to exploit the knowledge that $\Psi_1 = \hat{\Phi}_1 + \upepsilon$. Since we estimate $k$ single regression models, one has to use a different restriction for each estimation. For estimation $j$  note the a prioi information as
\begin{align*}
\underbrace{\begin{bmatrix}
\rm I_k 	& 	\rm O	\\
\end{bmatrix}}_{\bf R}
\uppsi_j 
=
\underbrace{
\begin{bmatrix}
\rm I_k 	& 	\rm O	\\
\end{bmatrix}
\upphi_j
}_{\rm {\bf r}_j}
+ 
\upepsilon_j
\end{align*} 
where $\rm O$ is a matrix of zeros in $ \mathbb{R}^{k \times k\cdot(p-1)}$. This allows to note the $f$-class estimator for estimation $j$ as
\begin{align*}
\rm 
\hat{\uppsi}^r_j = -
\Bigg(\frac{1}{\sigma^2}{\bf \Phi}'{\bf \Phi} + R' \Sigma(\Phi_1)_j^{-1}R\Bigg)^{-1}
\Bigg(\frac{1}{\sigma^2}{\bf \Phi}'\upphi_j + R' \Sigma(\Phi_1)_j^{-1} 
\big(-{\bf r}_j\big)\Bigg)
\end{align*}
Where $\Sigma(\Phi_1)_j$ is a matrix of zeros with the variances of the OLS parameter estimates for the $j$th column of $\Phi_1$ on its main diagonal. There is no covariance as long as one uses the standard OLS estimator for VARs, as described e.g.\ in \citet[][pp 71]{Luet}. This estimator  can be obtained by rewriting the VAR and transforming it into a standard linear regression model with the help of the Vec operator. As pointed out by \citeauthor{Luet}, this is identical to an OLS estimation of the $k$ equations separately.  This means that the estimation does not provide information about the covariance between the elements in the columns of $\Phi_i$, we therefore only use the variances. Similar to the univariate case, the variances of the MA estimates can be computed using just the left part of the estimator. 

\section{Conclusion}
This paper provides a simple method to estimate both univariate and multivariate MA processes by exploiting the recursive relation between the MA process and its AR representation. A simulation study for the parameter space of an MA(2) process indicates that the method tends to have a smaller MSE than Durbin's method and is relatively robust with respect to unit roots of the MA process, in the sense that the MSE increases moderately outside the invertibility triangle.

\bibliographystyle{plainnat}

\end{document}